\documentstyle[prd,aps,twocolumn]{revtex}
\newcommand{\al}{\alpha}

\newcommand{\ice}[1]{\relax}
\newcommand{\be}{\begin{equation}}
\newcommand{\ee}{\end{equation}}
\newcommand{\ba}{\begin{eqnarray}}
\newcommand{\ea}{\end{eqnarray}}
\newcommand{\nn}{\nonumber}
\newcommand{\MSsch}{{\overline{\rm MS}}}

\begin{document}
\draft
\preprint{MZ-TH/00-03}
\title{Strong coupling constant from
$\tau$ decay within a
renormalization scheme invariant treatment}
\author{ J.G.~K\"orner$^1$, F.~Krajewski$^1$, A.A.~Pivovarov$^{1,2}$}
\address{$^1$ Institut f\"ur Physik, Johannes-Gutenberg-Universit\"at,\\
  Staudinger Weg 7, D-55099 Mainz, Germany\\[12pt]
  $^2$ Institute for Nuclear Research of the
  Russian Academy of Sciences, Moscow 117312}
\maketitle
\begin{abstract}
We extract a numerical value for the strong coupling constant
$\alpha_s$ from
the $\tau$-lepton decay rate into nonstrange particles.
A new feature of our procedure is the explicit use of 
renormalization scheme invariance in analytical form 
in order to perform the actual analysis in a particular renormalization
scheme. For the reference coupling constant in the $\MSsch$-scheme
we obtain $\alpha_s(M_\tau)= 0.3184 \pm 0.0060_{exp}$ 
which corresponds to 
$\al_s(M_Z)= 0.1184 \pm 0.0007_{exp} \pm 0.0006_{hq\; mass}$.
This new numerical value is smaller than the standard value 
from $\tau$-data quoted in the literature 
and is closer to $\al_s(M_Z)$-values obtained 
from high energy experiments.
\end{abstract}

The physics of
$\tau$-lepton hadronic decays is 
an important area of particle phenomenology 
where the theory of strong interaction (QCD) can be confronted with experiment 
to a very high precision.
The central quantity of interest in this process is 
the
spectral density of hadronic states
related to 
the two-point correlator of
hadronic currents with well established and simple analytic properties.
The accuracy
of experimental data for a variety of observables 
of the $\tau$-lepton system is rather good and is steadily improving
\cite{exp1,exp2,PDG}. 
The spectral density itself 
(more precisely, the two-point correlator of
hadronic currents in the Euclidean domain) has been 
calculated with a very high
degree of accuracy within perturbation theory 
(see e.g. \cite{phys_report,eek20,eek21,eek2c}). 
Nonperturbative corrections to the correlator are known to 
be small and under control within the operator product expansion and 
factorization approximation \cite{SVZ,factor}.
The observables in the $\tau$ system are inclusive in nature which 
makes the comparison of experimental data with
theoretical calculations very clean
\cite{SchTra84,Bra88,Bra89,NarPic88,Pivtau,BraNarPic92,DP}.
Of some particular interest is the precise determination of the 
numerical value of the strong
coupling constant at the low energy scale of the $\tau$-lepton 
mass. Within the renormalization group approach 
this number can then be evolved to high energies.
This is a powerful consistency check of QCD 
since one is comparing hadron physics at 
a tremendous variety of scales, from one
to hundreds of ${\rm GeV}$ (e.g. \cite{wilczek}).

In the present note we provide a thorough analysis 
of the procedure of extracting numerical values of $\al_s$
from $\tau$-data in 
perturbation theory. 
On the theory side one expects a high degree of accuracy in 
the determination of $\al_s$ because of 
the existence of very accurate perturbation theory formulas 
and the simplicity of the renormalization group treatment 
of the massless quark case.
However, the numerical value of
the expansion
parameter $\al_s$ is not small at the $M_\tau$ scale
and the contribution of higher order terms in the 
perturbation theory series 
can be significant.
Arguments have been brought forth that the accuracy 
of finite-order perturbation theory is already close 
to its asymptotic limit
which makes the interpretation (usually called resummation) of 
the perturbation theory series in higher orders necessary \cite{tau12}.
The resummation of contributions related to the running of the
coupling constant is most advanced e.g. 
\cite{qed,be01,be02,be03,renZakh,bigi2,renRS}.

The decisive new point of our analysis 
is the explicit use of renormalization group invariance
in the analysis of the 
$\tau$-lepton decay rate within perturbation theory.
Renormalization group invariance is a fundamental property
of perturbation theory in quantum field theory 
which is related to the freedom 
in defining the subtraction procedure \cite{RG}. 
It should be respected in any numerical analysis.
Renormalization group invariance allows one to formally 
perform the numerical analysis in any renormalization scheme 
because all schemes are connected by a renormalization group
transformation. 
However, in the finite-order perturbation theory approach
this equivalence is only approximate 
due to the systematic omission of higher order terms 
in the perturbation theory 
expressions. This inroduces numerical differences into the results
obtained in different renormalization schemes.
Generally one can consider two ways of using perturbation theory
calculations.
One is to find relations between physical observables 
which are renormalization group invariant. 
Then perturbation theory calculations are just 
a purely intermediate step for finding
relations between observables (see, e.g. 
\cite{tau12,prl}) and no numerical analysis 
for renormalization scheme noninvariant quantities is performed. 
Indeed, let the perturbation theory expressions 
for two observables ${\cal O}_{1,2}$
in a given scheme have the form
\ba
\label{twoO}
{\cal O}_1&=&\al_s+r_1\al_s^2+O(\al_s^3), \nn \\
{\cal O}_2&=&\al_s+r_2\al_s^2+O(\al_s^3) \, .
\ea
Then the perturbation theory 
relation between observables ${\cal O}_{1,2}$ reads 
\be
\label{reltwoO}
{\cal O}_2={\cal O}_1+(r_2-r_1){\cal O}_1^2+O({\cal O}_1^3)
\ee
and is scheme-independent. 
The difference $r_2-r_1$ takes the same value for calculations in any scheme.
Another way of using perturbation theory
calculations is to extract numerical values for 
renormalization scheme noninvariant
quantities (as the coupling constant in a fixed scheme).
These are then compared with the results of other experiments. 
In this case the truncation of the perturbation theory series 
leads to numerical
violations of renormalization scheme invariance 
and plays an essential role.
In our simple example this means that the relations in eq.~(\ref{twoO})
are treated as quadratic functions of $\al_s$ in some fixed scheme and the
accuracy of extraction of the coupling constant value (and prediction
of other observables) depends drastically on the scheme
used, i.e. on the numerical values of the coefficients $r_{1,2}$. 

In the present paper we consider just this second application
and extract a numerical value
for the coupling constant which is not an immediate physical quantity.
By convention the reference value of the coupling constant
that is used to compare between
different experiments is fixed to be the $\MSsch$-scheme one.
However, and this is our point in this paper,
this does not necessarily
mean that for its extraction from a given experiment 
the numerical analysis 
should be performed in the $\MSsch$-scheme. 
It can be more convenient (and numerically accurate) to
analyze the system in its internal scheme and after finding numerical
values for the internal parameters translate them into 
the $\MSsch$-scheme using
renormalization scheme transformation. 
This program heavily uses explicit renormalization scheme covariance of
the theory. 
However, expressions for the amplitudes are available
only in perturbation theory as a truncated series in the coupling constant. 
For a truncated series the renormalization scheme
invariance is only approximate with a precision of the order of
the value of the first omitted term. Therefore numerical values 
obtained in the $\MSsch$-scheme
directly and through renormalization group transformations can differ. 
We discuss this
problem and argue that the internal scheme results are most reliable
physically and are more stable numerically
than the results of the standard analysis in the $\MSsch$-scheme.
Then numerical values for the reference $\MSsch$-scheme parameters can be 
obtained by a renormalization group 
``rotation'' from the numerical values found in the internal schemes. 
Renormalization group ``rotation'' (the re-calculation 
of numerical values from one scheme to another) 
is a quite formal operation and can be easily controlled
numerically. One example of such a ``rotation'' 
(the renormalization group scaling which is 
a one-parameter subgroup of the general renormalization group) 
is the evolution of the coupling
constant to the reference scale $M_Z$.
Below we give a detailed description of our approach.

The normalized 
$\tau$-lepton decay rate into nonstrange hadrons $h_{S = 0}$ 
is given by 
\ba 
\label{rate}
R_{\tau S=0}&=&\frac{\Gamma(\tau \rightarrow h_{S=0} \nu)}
{\Gamma(\tau \rightarrow l \bar{\nu} \nu )} \nn \\
&=& N_c |V_{ud}|^2 S_{EW}(1+\delta_P + \delta_{EW} + \delta_{NP})
\ea
where $N_c=3$ is the number of colors.  
The first term in eq.~(\ref{rate}) is the parton model 
result while the second term 
$\delta_P$ represents perturbative QCD
effects.
For the flavor mixing matrix element 
we use $|V_{ud}|^2 = 0.9511 \pm 0.0014$ \cite{PDG}.
The factor $S_{EW} = 1.0194 $ is an electroweak correction term 
\cite{ewcorr1} and $\delta_{EW} = 0.001$
is an additive electroweak correction \cite{ewcorr2}.
The nonperturbative corrections are rather small
and consistent with zero;
we use 
$\delta_{NP} = -0.003 \pm 0.003$ (see e.g. \cite{NarPic88}).
Note that recently the problem of duality violation 
for two-point correlators has been discussed \cite{chib,bigi}.
However, no established quantitative estimates
of that violation are available yet.
Considerations show that they can be rather large 
and can reach the level of few percents. 
This problem can affect the numerical value of the coupling extracted 
from the analysis because of the numerical change of the quantity 
$\delta_{P}$ extracted from eq.~(\ref{rate}).
In the present note we concentrate on
the perturbative part of the decay 
rate and numerical uncertainties related to 
the renormalization scheme
freedom of perturbation theory. 
In this respect new possible corrections do not qualitatively affect 
our analysis. The corrections due to duality violation
are of a new nature and they can be added independently to
eq.~(\ref{rate}). They would only change the input 
numerical value for the $\delta_{P}$ within our approach. 

The value for the decay rate
$R_{\tau S=0}$ has been measured by the ALEPH \cite{exp1} and OPAL \cite{exp2} 
collaborations with results very close
to each other.
For definiteness we use the ALEPH data
and briefly comment on the
OPAL data later on. With the experimental result
\be
\label{expdec}
R_{\tau S=0 }^{exp}=3.492 \pm 0.016
\ee
one obtains from eq.~(\ref{rate})
\be
\label{expdec0}
\delta_{P}^{exp}=0.203\pm0.007 \ .
\ee
The basic object of the theoretical calculation is 
Adler's
$D$-function which is computable in perturbation theory in 
the Euclidean
domain.
In the $\overline{\rm MS}$-scheme the perturbative expansion for the 
$D$-function is given by 
\ba
\label{spect0}
D(Q^2)&=&1+\frac{\al_s(Q)}{\pi}
+k_1 \left(\frac{\al_s(Q)}{\pi}\right)^2 
+k_2 \left(\frac{\al_s(Q)}{\pi}\right)^3 \nn \\ 
&&
+ k_3 \left(\frac{\al_s(Q)}{\pi}\right)^4+O(\al_s(Q)^5)
\ea
with (see e.g. \cite{phys_report})
\ba
  \label{ks}
k_1&=&\frac{299}{24} - 9\zeta(3), \nn \\
k_2&=&\frac{58057}{288} - \frac{779}{4}\zeta(3) +
\frac{75}{2}\zeta(5)\, .
\ea
Here $\zeta(x)$ is Riemann's $\zeta$-function.
In the following we use the notation 
\be 
a_s(Q)=\frac{\al_s(Q)}{\pi}
\ee
for the standard $\MSsch$-coupling constant
normalized at the scale 
$\mu=Q$. 
Numerically we find
\ba
\label{spect}
D(Q^2)&=&1+a_s(Q)+1.6398 a_s(Q)^2 + 6.3710 a_s(Q)^3 \nn \\ 
&&+ k_3a_s(Q)^4 +O(a_s^5(Q))\, .
\ea
The coefficient $k_3$ is still unknown
which prevents us from using the last term in eq.~(\ref{spect})
for our analysis.
We nevertheless list this term throughout the paper to 
obtain a feeling for the possible 
magnitude of the $O(\al_s^4)$ correction.
The particular numerical value of $k_3\sim 25$ is 
obtained on the basis of geometric series approximation for the series
(\ref{spect}) and is often used in the literature 
\cite{DP,kataev,Elias}.
In our analysis we do not use any particular numerical value for 
$k_3$ and only give some illustrative
results of the influence of this term on 
the numerical value of the coupling constant 
extracted from $\tau$-data.

In the $\MSsch$-scheme
the perturbative correction $\delta_{P}$
is given by the perturbation theory expansion
\ba
\label{taumssch}
\delta_P^{th} &=& a_s + 5.2023 a_s^2
+26.366a_s^3 \nn \\
&& +(78.003+k_3)a_s^4 + O(a_s^5)
\ea
where the $\MSsch$-scheme coupling constant 
$\al_s=\pi a_s$ is taken at 
the scale of the $\tau$-lepton mass
$\mu=M_\tau = 1.777~{\rm GeV}$.
Usually one extracts a numerical value for $\al_s(M_\tau)$
by treating
the first three terms of the expression in eq.~(\ref{taumssch})
as an exact function -- the cubic polynomial, i.e. one solves
the equation 
\ba
\label{taumsschEx}
a_s + 5.2023 a_s^2
+26.366a_s^3=\delta_P^{exp}\, .
\ea
The solution reads
\be
\label{dirres}
\pi a_s^{st}(M_\tau)\equiv \al_s^{st}(M_\tau)= 0.3404\pm 0.0073_{exp} \, .
\ee
We call this method the standard method.
The quoted error is due to the error in the 
input value of $\delta_P^{exp}$.
We retain some additional decimal points in the 
numerical expression for the coupling constant in order 
to use them 
for the evolution of the coupling constant to the scale $M_Z$. 
It is rather difficult to estimate the theoretical uncertainty of the
procedure itself. The main problem is to estimate the quality 
of the approximation for the (asymptotic)
series in eq.~(\ref{taumssch}) given by
the cubic polynomial in eq.~(\ref{taumsschEx}).

As a criterion of the quality of the approximation 
one can use the pattern of convergence of the series 
(\ref{taumssch}) which is 
\be
\delta_P^{exp}=0.203=0.108+0.061+0.034+\ldots
\ee
One sees that the corrections provide a 
100\% change of the leading term.
Another criterion is the order-by-order behavior of the extracted
numerical value for the coupling constant. In consecutive orders
of perturbation theory (LO - leading order, NLO - next-to-leading
order, NNLO - next-next-to-leading order)
one has
\ba 
&&\al_s^{st}(M_\tau)_{LO}=0.6377,\quad 
\al_s^{st}(M_\tau)_{NLO}=0.3882, \nn \\ 
&&\al_s^{st}(M_\tau)_{NNLO}=0.3404\, .
\ea
Formally we obtain a series for the numerical value of the 
coupling constant of the form 
\ba
\al_s^{st}(M_\tau)_{NNLO}&=&0.6377-0.2495-0.0478-\ldots
\ea
Limiting ourselves to the next-to-next-to-leading order
result (NNLO) we can take a half of the last term
as an estimate of the theoretical 
uncertainty. It is only an indicative estimate.
No rigorous justification can be given for such an assumption
about the accuracy of the approximation without knowledge
of the structure of the whole series.
Nevertheless we stick to this definition for our purposes.
The theoretical uncertainty obtained in 
such a way -- 
$\Delta \al_s^{st}(M_\tau)_{th}=0.0239$ -- 
is much larger than the experimental uncertainty
given in eq.~(\ref{dirres}).
This is a challenge for the theory: the accuracy of 
theoretical formulas cannot
compete with experimental precision at present. 
Assuming this theoretical uncertainty we have 
\be
\label{finalST} 
\al_s^{st}(M_\tau)_{NNLO}=0.3404\pm 0.0239_{th}\pm 0.0073_{exp}.
\ee
Theory dominates the error even if the estimate for 
its precision
$\pm 0.0239_{th}$ is not reliable (heuristic
and only indicative).
Thus the straightforward analysis in the $\MSsch$-scheme
is not stable numerically and the naive estimate of the theoretical
uncertainty is large.

The use of the $\MSsch$-scheme is not obligatory for 
practical calculations.
The $\MSsch$-scheme has a history of success for
massless calculations where its results look natural and 
the corrections are usually small. This is not the strict rule, however, 
and there are cases (like gluonic correlators \cite{glu})
where corrections dramatically depend on the quantum numbers
of the operators. 
In fact, the $\MSsch$-scheme is rather artificial.
It is simply defined by convention
(let us be remindful of 
the evolution from the MS-scheme to the $\MSsch$-scheme which had
its origin only in technical convenience \cite{buras}).
From technical point of view,
in practical calculations of massless diagrams of the propagator type, 
another scheme -- the $G$-scheme -- 
is the most natural one \cite{Gsch}. It normalizes the basic quantity
of the whole calculation within 
integration-by-parts technique -- one loop masless 
scalar diagram -- to unity \cite{intbyparts}.
$\beta$-functions coincide in both schemes.
It could have well happened that the $G$-scheme would be historically 
adopted as the reference scheme
because corrections in this scheme are typically smaller than that in 
the $\MSsch$-scheme.
However, for the tau system the direct (standard) 
analysis in the $G$-scheme fails.

Therefore different schemes used for the numerical analysis
can produce rather different numerical results for the final
reference quantity - the coupling constant in the $\MSsch$-scheme.
Note that strictly speaking any scheme is suitable for 
a given perturbative calculation.
However, it can lead to unusual (or
even unacceptable) results in a numerical analysis. 
The only criterion for the choice of scheme at present is the
heuristic requirement of fast explicitl convergence: the terms of the 
series should decrease. Clearly this is a rather unreliable criterion.
It does not provide strict quantitative constraints necessary for the
level of precision usually claimed for the $\tau$-system analysis.

In the following we suggest a new procedure for extracting 
$\al_s$ in the $\MSsch$-scheme from the $\tau$ system 
without explicit use 
of eq.~(\ref{taumssch}).
This procedure is applicable to any observable in whatever scheme
it was originally computed.
The observation is that any perturbation theory observable 
generates a scale due to
dimensional transmutation and this is its internal scale.
It is natural for a numerical analysis (and is our suggestion) 
to determine this scale fisrt and then to transform 
the result into 
a $\MSsch$-scheme
parameter (or any other reference scheme) 
using the renormalization group invariance. 
We deliberately use the explicit renormalization scheme invariance 
of the theory to bring the result
of the perturbation theory calculation 
into a special scheme first, then we perform a
numerical analysis in this
particular scheme. Only after that 
we transform the obtained
numbers into 
the reference $\MSsch$-scheme. The last step is done only for comparison
with other experiments (or just for convenience; 
the system itself can be well described in its internal scheme 
without any reference to the $\MSsch$-scheme). 
This is our suggestion for 
the resolution of the problem of numerical instability of extracting
parameters from truncated perturbation theory expressions.

A dimensional scale in QCD emerges 
as a boundary value parameterizing the evolution
trajectoty
of the coupling constant.
The renormalization group equation 
\be \label{RGE}
\mu^2 \frac{d}{d \mu^2} a(\mu^2) = \beta(a(\mu^2))\, ,
\quad a=\frac{\al}{\pi} 
\ee
is solved by the integral
\be 
\label{LQCD}
\ln \left(\frac{\mu^2}{\Lambda^2}\right) =
   \Phi(a(\mu^2)) + \int_0^{a(\mu^2)} 
\left( \frac{1}{\beta(\xi)} - \frac{1}{\beta_2(\xi)} \right)d\xi
\ee
where the indefinite integral $\Phi(a)$ is normalized 
as follows
\be 
\label{intb2}
\Phi(a) = \int^{a}  \frac{1}{\beta_2(\xi)} d \xi 
= \frac{1}{a \beta_0} 
       + \frac{\beta_1}{\beta_0^2} 
\ln \left( \frac{a \beta_0^2}{\beta_0 + a \beta_1} \right).
\ee
Here
$\beta_2(a)$ and $\beta(a)$ denote the second order 
and full $\beta$ function, or as many terms as are available,
given by
\ba
\beta_2(a)&=& -a^2(\beta_0 + a \beta_1), \nn \\
\beta(a)&=& -a^2(\beta_0 + \beta_1 a + \beta_2 a^2 
+ \beta_3a^3) +O(a^6)\, ,
\ea
$a$ is a generic coupling constant.
The four-loop $\beta$-function coefficient $\beta_3$ is now known in 
the $\MSsch$-scheme \cite{beta4}
\be
\beta_3=\frac{140599}{4608} + \frac{445}{32}\zeta(3)=47.228\ldots
\ee
The integration constant in eq.~(\ref{LQCD}) 
is adjusted such that the asymptotic expansion 
of the coupling constant at large momenta $Q^2\to \infty$ reads
\ba 
\label{Lser}
a(Q^2) &=&  \frac{1}{\beta_0 L}\left(1 - \frac{\beta_1}{\beta_0^2} 
\frac{\ln(L)}{L^2} \right)+ O \left(\frac{1}{L^3} \right)
\ , \nn \\
L&=&\ln\left(\frac{Q^2}{\Lambda^2}\right) \, . 
\ea
This serves to define the parameter $\Lambda$ 
(dimensional scale) for a generic coupling
constant.
$\Lambda_s$ is the standard $\MSsch$-scheme 
scale for the coupling constant $a_s$.

The solution (\ref{LQCD}) of the renormalization group equation (\ref{RGE}) 
describes the evolution trajectory of the coupling constant.
This trajectory is parametrized by the scale parameter 
$\Lambda$ and the coefficients of the $\beta$ function 
$\beta_i$ with $i>2$ (see e.g. \cite{stevenson}). The evolution is 
invariant under the renormalization group transformation
\be 
\label{kappa}
a  \rightarrow a(1 + \kappa_1 a + \kappa_2 a^2 + \kappa_3 a^3 + \dots )
\ee
with the simultaneous change
\be
\label{transform}
\Lambda^2 \rightarrow  \Lambda^2 e^{-\kappa_1/\beta_0}\, ,
\ee
$\beta_{0,1}$ left invariant and
\ba
\beta_2 &\rightarrow&\beta_2-\kappa_1^2\beta_0
+\kappa_2 \beta_0 - \kappa_1 \beta_1 \nn \\
\beta_3 &\rightarrow&\beta_3+4 \kappa_1^3 \beta_0 
+ 2 \kappa_3 \beta_0 + \kappa_1^2  
\beta_1 - 2 \kappa_1 (3 \kappa_2 \beta_0 + \beta_2 )  .\nn
\ea
If this transformation was considered to be exact
and the exact $\beta$-function
corresponding to the new charge was used then it would be just a 
change of variable in a differential equation (\ref{RGE}) or
the exact reparametrization of the trajectory (\ref{LQCD})
and hence would lead to identical results.
However, the renormalization group invariance of eq.~(\ref{LQCD}) 
is violated in higher orders
of the coupling constant because we consistently omit higher orders 
in the perturbation theory expressions for 
the $\beta$-functions.
This is the point where the finite-order perturbation 
theory approximation for the respective $\beta$-functions
is made.
This is the source for different numerical outputs of analyses in
different schemes.

Our procedure for the extraction of $\al_s$ is
heavily based on 
the formal renormalization group invariance of the theory.
We claim that because of this invariance we can do our numerical analysis 
in any scheme. The reason for the choice of a particular scheme
is only the quality of the 
convergence (which, of course, is subject to some personal taste).
We have chosen the effective scheme because we consider it 
to be more consistent and more stable numerically. 

Technically we introduce an effective charge $a_\tau$ through the relation
\cite{prl,grunberg,krasK,dhar,brodsky}
\be
\delta_P^{th} = a_\tau\equiv \frac{\al_\tau}{\pi}
\ee
and extract the parameter $\Lambda_\tau$ which 
is associated with $a_\tau$ through eq.~(\ref{LQCD}).
This is just the internal scale associated with 
the physical observable $R_\tau$.
The effective $\beta$-function is given by the expression 
\be
\beta_\tau = -a_\tau^2(\beta_{\tau 0} + \beta_{\tau 1} a_\tau +
\beta_{\tau 2} a_\tau^2 + \beta_{\tau 3} a_\tau^3 + \dots) 
\ee  
with 
$\beta_{\tau 0} = \beta_0$, $\beta_{\tau 1 } = \beta_1$, and 
\be 
\label{effbeta}
\beta_{\tau 2 }= -12.3204\, ,
\quad
\beta_{\tau 3 }= -182.719 + \frac{9}{2} k_3 \, .
\ee
The extraction of the numerical value for the internal scale 
$\Lambda_\tau$ is done 
from equation (\ref{LQCD}) with 
$a_\tau(M_\tau) = \delta_P^{exp}$.
The coefficient $\beta_{\tau 3}$ does not enter the analysis.
The parameter $\Lambda_s\equiv \Lambda_{\MSsch}$ is found
according to eq.~(\ref{transform}).
The $\MSsch$ coupling at $\mu = M_\tau$ 
is obtained by solving eq.~(\ref{LQCD})
for $a_s(M_\tau)$ with regard to $\ln(M_\tau^2 / \Lambda_s^2)$ 
which is known if $\Lambda_s$ is obtained; the
$\beta$-function is taken in the $\MSsch$-scheme.
For consistency reasons we only use the $\MSsch$-scheme 
$\beta$-function 
to three-loop order since the effective $\beta$-function $\beta_\tau$
is only known up to the second order, cf. eq. (\ref{effbeta}).
A $N^3LO$ analysis is possible only if a definite value 
is chosen for $k_3$.
We give some estimates later.

Our procedure is based on 
renormalization group invariance and one can start from the
expression for the decay rate obtained in any scheme. 
The only perturbative objects present are the $\beta$-functions.
Both $\beta_{\MSsch}$ \ and $\beta_\tau$,
however, converge reasonably 
well which is the only perturbation theory restriction 
in our method. 
It also highlights the limit of precision within our procedure:
the expansion for $\beta_\tau$ is believed to be asymptotic 
as any expansion in
perturbation theory.
The asymptotic expansion provides only limited accuracy
for any given numerical value of the expansion parameter 
which cannot be further improved by including higher order terms.
The expansion used is 
presumably rather
close to its asymptotic limit
as can be seen by taking a look at the expansion
\ba
\label{betatau}
\beta_\tau(a_\tau)&=&- a_\tau^2 \Big( \frac{9}{4} + 4 a_\tau 
- 12.3204 a_\tau^2 \nn \\
&& + a_\tau^3 \left(-182.719+\frac{9}{2} k_3 \right) \Big) 
+O(a_\tau^6) 
\ea
with $a_\tau\sim 0.2$ at the scale $M_\tau$.
The convergence of the
series depends crucially on the numerical value of $k_3$.
If $k_3$ had a value where 
the asymptotic growth starts at third order
then further improvement of the accuracy within 
finite-order perturbation theory
is impossible.

At every order of the analysis we use the whole information of 
the perturbation theory
calculation.
Especially, the appropriate coefficient of the $\beta_\tau$-function 
is present.
In the standard method the coefficient $\beta_2$
enters only at order $O(\al_s^4)$ of the
$\tau$-lepton decay rate expansion.
We call our procedure the renormalization scheme invariant extraction 
method (RSI) hoping that it is clear what is meant by this name from
our explanations. Note also that $\al_s$ itself is not 
a physical
object and is renormalization scheme noninvariant. 
In this respect we extract the noninvariant
parameter $\al_s$ using invariance of the physics in order to perform
the numerical analysis in the most suitable scheme. 
Then the output of the
analysis is simply transformed into a numerical value for $\al_s$ 
according to the renormalization group transformation rules.
For the coupling constant in the $\MSsch$-scheme in NNLO
we find 
\ba
\label{finFO}
\al_{s}^{RSI}(M_\tau) &=& 0.3184 \pm 0.0060_{exp}  
\ea
which is smaller than the corresponding value obtained within 
the standard procedure eq.~(\ref{dirres}).
How to estimate the quality of this result?
The parameter which is really extracted in consecutive orders of perturbation theory 
within our method is
the scale $\Lambda_\tau$. Because of the relation
(see eqs. (\ref{taumssch},\ref{kappa},\ref{transform}))
\be
\Lambda_s=\Lambda_\tau e^{-5.20232/2\beta_0}=0.3147\Lambda_\tau
\ee
we can look at $\Lambda_s$ directly.
We find
\ba
&&\Lambda_s|_{LO}=595~{\rm MeV},\quad 
\Lambda_s|_{NLO}=288~{\rm MeV}, \nn \\
&&\Lambda_s|_{NNLO}=349~{\rm MeV}
\ea
or, representing the NNLO result as a formal series, 
\be
\label{lamser}
\Lambda_s|_{NNLO}=595-307+61-\ldots~{\rm MeV}.
\ee
Note that at leading order the scales (as well as charges) 
are equal in all schemes. Therefore the leading order result 
($\Lambda_s|_{LO}=595~{\rm MeV}$) is not representative,
only indicative.
Assuming according to our convention that the uncertainty of $\Lambda_s$
is given by the half of the last term of the series (\ref{lamser}) we have
\be
\Lambda_s=349\pm 31~{\rm MeV}
\ee
which leads to the numerical value for the $\MSsch$-scheme coupling
constant 
\be
\al_s=0.3184^{-0.0157}_{+0.0160} \, .
\ee
This result is obtained from eq.~(\ref{LQCD}) with three-loop
$\beta$-function. Taking the average we find 
\be
\label{theoryNNLO}
\al_s=0.3184 \pm 0.0159\, .
\ee
This is better than the theoretical error of the standard result 
eq.~(\ref{finalST}). 
Still the theoretical error should be considered as a guess 
rather than a well-justified estimate of the uncertainty.

Let us briefly comment on the $k_3$ contribution.
Clearly the estimate $k_3=25$ is rather speculative.
We, therefore, use a different strategy
in the analysis. We determine the range of $k_3$
which is safe for explicit convergence 
of perturbation theory. If the actual value of $k_3$ will be 
discovered in this range 
then perturbation theory is still valid 
and will give better accuracy in NNNLO. 
If not, the asymptotic growth of perturbation theory series is 
already reached and its accuracy cannot be improved.

We require 
that the last term is equal to the half of the previous one. 
In the standard way (eq.~\ref{taumssch}) we have 
\be
|(78+k_3)a_s|<\frac{1}{2}\, 26.36 \approx 13
\ee
which for $a_s=0.1$ gives
\be
\label{standrest}
-208<k_3^{st}<52 \, .
\ee 
In the RSI way (eq.~\ref{betatau}) we have 
\be
|(-182+\frac{9}{2}k_3)a_\tau|<\frac{1}{2}\, 12.32 \approx 6
\ee
which for $a_\tau=0.2$ gives
\be
33.8<k_3^{\tau}<47.1 \, .
\ee 
This range is much narrower 
than that in eq.~(\ref{standrest}).
The effective scheme method is much more sensitive 
to the structure of the series as can be seen from eq.~(\ref{betatau}).
The 
actual precision depends on the actual value chosen for $k_3$
and it is rather premature to speculate about 
numbers.

Still we show the worst result 
(in the optimistic scenario that $k_3$ lies in the safe
range)
that can be expected within the RSI approach.
In the RSI approach with $k_3=47$ we find the scale parameter in NNNLO
\be
\Lambda_s|_{NNNLO}=334~{\rm MeV} \, .
\ee
With $k_3=34$ one has 
\be
\Lambda_s|_{NNNLO}=367~{\rm MeV}  \, .
\ee
Taking the average we have 
\be
\Lambda_s=350\pm 17~{\rm MeV}
\ee
which is the best possible
estimate if we require that the perturbation theory series for 
the $\beta_\tau$-function still 
converges (according to our quantitative criterion of convergence).
That results in the numerical value for the $\MSsch$-scheme coupling
constant found with four-loop $\beta$-function from eq.~(\ref{LQCD})
\be
0.3133< \al_s< 0.3314 \, .
\ee
Therefore our conservative estimate of the theoretical error
in the optimistic scenario
for the convergence of perturbation theory series in NNNLO
reads
\be
\label{optNNNLO}
\al_s=0.322\pm 0.009\, .
\ee 
While the estimation of the theoretical uncertainty is a tricky matter
and can be considered as indicative 
the central numerical value of the coupling constant 
definitely becomes smaller as compared to the standard result. 

At present the reference value for the coupling constant is
commonly given
at the scale $M_Z = 91.187~{\rm GeV}$. 
The running to this reference scale
is done with the four-loop $\beta$-function in 
the $\MSsch$-scheme \cite{beta4} 
and three-loop matching conditions
at the heavy quark (charm and bottom)
thresholds \cite{matching}. 
For the threshold parameters related to heavy quark masses 
we use 
$\mu_c=\bar{m}_c(\mu_c)=(1.35\pm 0.15)~{\rm GeV}$
and
$\mu_b=\bar{m}_b(\mu_b)=(4.21\pm 0.11)~{\rm GeV}$ (e.g. \cite{bbmass})
where $\bar{m}_q(\mu)$ is the running mass of the heavy quark in 
the $\MSsch$-scheme.
Note that because of the truncation of matching conditions
the result of the running slightly depends on at what scale 
the matching is actually performed. 
If the matching between the $n_f=3$ 
and $n_f=4$ effective theories is done directly at the scale $M_\tau$,
which is possible, then the result is slighly smaller than in the case
when the evolution within $n_f=3$ effective theory is done first 
to the scale $\mu_c$. In the following we stick to the procedure 
where 
the matching is performed precisely at 
the matching scales $\mu_{c,b}$. 
We first run the coupling constant within $n_f=3$ effective theory
from the scale $M_\tau$ to $\mu_c$ then match the result to 
$n_f=4$ coupling constant, run it to $\mu_b$ and match to 
$n_f=5$ coupling constant. The last step is just evolution to $M_Z$.
Note that the alternative would be to perform matching between 
$n_f=3$ and $n_f=4$ effective theories directly at the scale 
$M_\tau$ (because it is rather close to $\mu_c$) but in this case the
final result is slightly smaller than in our present procedure.

The running to the scale $M_Z$ 
gives the following result for the standard method estimate
\be
\label{dir0}
\al_s^{st}(M_Z)=0.1210 \pm 0.0008_{exp} \pm 0.0006_c \pm 0.0001_b
\ee
where the subscript $exp$ denotes the error 
originating from $\delta_P^{exp}$.
The errors with subscripts 
$c,b$ arise from the uncertainty of the numerical values of 
the charm and bottom quark
masses that enter the evolution analysis. 
These errors are rather small
(we retain the additional decimal place in the result, 
which is not
really justified from the precision of the
experimental input, just to show these uncertainties).
If the matching between the $n_f=3$ 
and $n_f=4$ effective theories is done directly at the scale $M_\tau$
one has to change the central value $0.1210\to 0.1202$
which shows the uncertainty related to the truncation of 
the matching conditions. 

The central value in eq.~(\ref{dir0}) is slightly higher than
that calculated from high energy experiments \cite{PDG}.
The theoretical perturbative expansions
for observables in 
high energy experiments
converge better numerically 
than expansions at low energies
because the coupling, which is the parameter of 
the perturbative expansion,
is smaller at higher energies 
due to the property of asymptotic
freedom. 
This feature
makes it less important to treat  the 
higher order terms
carefully in high energy applications 
as compared to the low
energy $\tau$-lepton estimates.
However, the experimental data in high energy experiments 
are usually less precise which leads to large errors in
the $\al_s$ determination
from high energy experiments.
The fact that the value in eq.~(\ref{dir0})
is higher than that calculated from high energy experiments
caused some discussion about the reliability of estimates from the 
$\tau$-lepton data.
Our analysis resolves this problem.
The running of
$\al_s^{RSI}(M_\tau)$ 
given in eq.~(\ref{theoryNNLO})
to $M_Z$ with the four-loop $\beta$-function and 
with three-loop heavy quark matching
accuracy gives
\ba
\label{rgi0}
\al_{s}^{RSI}(M_Z) &=& 0.1184 \pm 0.00074_{exp} \nn \\
&&\pm 0.00053_c \pm 0.00005_b 
\ea
where we have kept five decimal places in order to exhibit
the magnitude 
of different sources of uncertainty.
Eq.~(\ref{rgi0}) constitutes our main result  
for the coupling $\al_{s}(M_Z)$ derived from tau data.

The OPAL collaboration has reported an experimental value
of $R_{\tau S=0}^{exp} = 3.484 \pm 0.024$ \cite{exp2}.
This leads to
$\delta_P^{exp} = 0.200 \pm 0.009_{exp}$
and
\be
\al_s^{RSI}(M_\tau) =0.3158 \pm 0.0078_{exp}
\ee
which, when evolved to $M_Z$, gives
\ba
\al_s^{RSI}(M_Z) 
&=&0.1181 \pm 0.00097_{exp} \nn \\
&&\pm 0.00052_c \pm 0.00005_b \, .
\ea
This value is close to the one in eq.~(\ref{rgi0}) based on 
the ALEPH data.

The theoretical uncertainty comes mainly from 
the truncation of the perturbation theory series.
Taking the result of the NNLO analysis eq.~(\ref{theoryNNLO})
we find 
\be
\Delta \al_s^{RSI}(M_Z)_{th}=0.0019
\ee 
In the most optimistic scenario with the NNNLO
analysis eq.~(\ref{optNNNLO}) one has 
\be
\al_s^{RSI}(M_Z)_{N^3LO}=0.119\pm0.001\, .
\ee

As we have already noted the interpretation of 
the higher order terms 
in the perturbation theory expansion is numerically important for the
analysis of the $\tau$-data.
The regular method to resum higher order perturbation theory
corrections is
based on the direct integration of the renormalization group improved 
correlators over the contour in the
complex $Q^2$ plane \cite{Pivtau}.
This method allows one to resum corrections generated by the running 
of the coupling constant along the integration contour
and is now widely used for the analysis of the $\tau$-data.
We now briefly comment on the extraction of 
the strong coupling constant within resummed 
perturbation theory.
As in ref.~\cite{resana}
we fit the theoretical expression for the 
decay rate in the contour improved approach
to the experimental 
result $\delta_{P}^{exp}$ eq.~(\ref{expdec0}) and find
\be 
\label{CI}
\al_s^{CI}(M_\tau) = 0.343 \pm 0.009_{exp}
\ee
within the renormalization scheme invariant extraction 
method described above i.e. with the introduction of 
the effective charge first. 
This value differs from the finite-order perturbation theory result
eq.~(\ref{finFO}). 
Note that the two values extracted from 
finite-order perturbation theory analysis eq.~(\ref{finFO})
and the contour improved perturbation theory analysis
eq.~(\ref{CI})
do not overlap within
their respective error bars
given from the experimental uncertainty only.
This situation was anticipated in \cite{Pivtau} 
where the resummed NNLO analysis had been
first performed.
The point is clear: resummation provides
a specific estimate of higher order terms.
In finite-order perturbation theory one adopts a model where all 
higher order terms have been neglected.
In contour improved perturbation theory one adopts an
explicit model with higher order
terms generated by the running of the coupling constant 
along the integration contour. With present 
experimental accuracy one can
already distinguish between these two possibilities. 
One should always
keep in mind that the two determinations eq.~(\ref{finFO})
and eq.~(\ref{CI}) result
from different models and one should not mix
their predictions. 

The numerical value of the coupling constant 
appropriate for high energy experiments
is normally small (much smaller than for $\tau$-data)
and perturbation theory converges faster 
(in similar kinematical situations).
The resummation does not produce any big numerical changes.
Therefore finite-order perturbation theory 
is normally used
for the analysis of high energy
experiments
(resummation of the contour type 
can be done but produces a small numerical effect) and 
one usually quotes numerical
values of the coupling constant extracted with 
finite-order perturbation theory.
Or resummation of the sort different from that used for 
the $\tau$ system is used 
(like Coulomb type resummation for heavy quarks \cite{vol,hoang}).
Therefore we suggest to use
the finite-order perturbation theory
prediction for the coupling constant extracted from $\tau$-data
in order to compare it with the results
of high energy experiments.

To conclude,
we have extracted the numerical value of the strong coupling constant
from $\tau$-data
within a procedure based on explicit use of 
renormalization scheme invariance.
The numerical value for the coupling constant 
is systematically smaller than that derived by the standard treatment.
When evolved to $M_Z$ our $\MSsch$-scheme value for the
coupling constant extracted in finite-order perturbation 
theory reads 
\be
\al_s(M_Z) = 0.1184 \pm 0.0007_{exp} \pm 0.0006_{hq\; mass} \, .
\label{finresMZ}
\ee
This central value is closer to the value of $\al_s$ derived from high
energy experiments
than previous determinations of $\al_s$ from $\tau$-data.
The theoretical uncertainty of the result is still only indicative:
it ranges from the conservative estimate in NNLO
$\Delta \al_s(M_Z)_{th}=\pm 0.0019$ to an optimistic one based on the  
assumption about NNNLO contribution
$\Delta \al_s(M_Z)_{th}=\pm 0.001$.

The present work is supported in part by the Volkswagen 
Foundation under contract No.~I/73611 and 
by the Russian Fund for Basic Research under contract
99-01-00091.

\end{document}